\documentclass[twocolumn,floatfix,superscriptaddress,a4paper,showkeys,nofootinbib,reprint]{revtex4-1}

\usepackage[colorlinks=true,linktocpage=true,linkcolor=blue,citecolor=blue,allcolors=blue]{hyperref}
\usepackage{epsfig}
\usepackage{latexsym}
\usepackage[utf8]{inputenc}
\usepackage{xspace}
\usepackage{indentfirst}
\usepackage{enumitem}
\usepackage{color}
\usepackage{setspace}
\usepackage{lipsum}
\usepackage{soul}
\usepackage{todonotes}

\usepackage{amsmath}
\usepackage{amssymb}
\usepackage[english]{babel}
\usepackage{url}

\newcommand{\eq}[1]{\begin{align} #1 \end{align}}

\begin{document}
\title{Enhanced dilepton emission from a phase transition in dense matter}
\author{Oleh Savchuk}
\affiliation{GSI Helmholtzzentrum f\"ur Schwerionenforschung GmbH, D-64291 Darmstadt, Germany}
\affiliation{Bogolyubov Institute for Theoretical Physics, 03680 Kyiv, Ukraine}
\affiliation{Frankfurt Institute for Advanced Studies, Giersch Science Center,
D-60438 Frankfurt am Main, Germany}
\author{Anton Motornenko}
\affiliation{Frankfurt Institute for Advanced Studies, Giersch Science Center,
D-60438 Frankfurt am Main, Germany}
\author{Jan Steinheimer}
\affiliation{Frankfurt Institute for Advanced Studies, Giersch Science Center,
D-60438 Frankfurt am Main, Germany}	
\author{Volodymyr~Vovchenko}
\affiliation{Institute for Nuclear Theory, University of Washington, Box 351550, Seattle, WA 98195, USA}
\affiliation{Nuclear Science Division, Lawrence Berkeley National Laboratory, 1 Cyclotron Road,  Berkeley, CA 94720, USA}
\affiliation{Frankfurt Institute for Advanced Studies, Giersch Science Center,
D-60438 Frankfurt am Main, Germany}
\author{Marcus Bleicher}
\affiliation{Goethe-Universität Frankfurt, Institut f\"ur Theoretische Physik, Max-von-Laue-Str. 1, 60438 Frankfurt am Main, Germany}
\affiliation{Helmholtz Research Academy Hesse (HFHF), GSI Helmholtzzentrum f\"ur Schwerionenforschung GmbH, Campus Frankfurt, 60438 Frankfurt am Main, Germany}
\author{Mark Gorenstein}
\affiliation{Frankfurt Institute for Advanced Studies, Giersch Science Center,
D-60438 Frankfurt am Main, Germany}
\affiliation{Bogolyubov Institute for Theoretical Physics, 03680 Kyiv, Ukraine}
\author{Tetyana Galatyuk}
\affiliation{GSI Helmholtzzentrum f\"ur Schwerionenforschung GmbH, D-64291 Darmstadt, Germany}
\affiliation{Technische Universit\"at Darmstadt, 64289 Darmstadt, Germany}

\date{\today}

\begin{abstract}
It is demonstrated that the presence of a phase transition in heavy ion collisions, at beam energies that probe dense QCD matter, leads to a significant enhancement of the dilepton yield per produced pion due to the extended emission time.
In addition, the temperature of low mass dileptons shows a modest decrease due to the mixed phase.
The emission of dileptons in the SIS18--SIS100 beam energies range is studied by augmenting the UrQMD transport model with a realistic density dependent equation of state, as well as two different phase transitions. This is achieved by extending the molecular dynamics interaction part of the UrQMD model to a density dependent interaction potential with a high density minimum leading to a phase transition and metastable coexisting high density states.
Together with a high precision measurement these simulations will be able to constrain the existence of a phase transition in QCD up to densities of several times nuclear saturation density.
\end{abstract}

\maketitle

\section{Introduction}
The high density structure of the QCD phase diagram is one of the major challenges in theoretical and experimental high energy physics. 
Relativistic heavy ion collisions~(HIC) probe different regimes of QCD matter by varying the collision energy and mass of colliding nuclei. Complementary, the low-temperature and high-density region of the phase diagram is probed by neutron stars and, since the detection of the GW170817 gravitational wave event and its electromagnetic counterpart~\cite{LIGOScientific:2017vwq,LIGOScientific:2020aai}, by their mergers.
At vanishing net-baryon density, first principle lattice QCD simulations show a smooth chiral crossover~\cite{Borsanyi:2013bia,HotQCD:2014kol,Bazavov:2017dus} which can be probed by the vast amount of data collected from HIC at highest energies, e.g. at the LHC and RHIC~\cite{Bzdak:2019pkr,ALICE:2019nbs,STAR:2021iop,STAR:2021rls}.
On the other hand, the properties of matter at high and even intermediate net-baryon densities are still mostly unconstrained and the existence of a first order phase transition, with its corresponding critical end point, in this region is not ruled out.
Besides the constraints from astrophysical observations of compact stars one can only rely on signatures of a phase transition in HIC.
Here, the challenge is to make unambiguous predictions for current and future experiments, which require to embed a first order phase transition into the hitherto existing models for the dynamical description of HIC. 

Several experimental signals were suggested as sensitive probes for the phase structure of the matter created in HIC.
For example, the collapse of the directed flow~\cite{Rischke:1995pe,Stoecker:2004qu,Brachmann:1999xt,Brachmann:1999mp}, fluctuations and correlations~\cite{Stephanov:1998dy,Stephanov:1999zu,Jeon:2000wg,Asakawa:2000wh,Bzdak:2019pkr}, as well as light nuclei enhancement~\cite{Sun:2017xrx}, just to name a few.
The experimental confirmation of these signatures is challenging, and no final conclusion has been reached so far.
It is important to note that such signals cannot be interpreted isolated from the wealth of data which is now available and in the end only a fully consistent combination of observables can lead to conclusions on the phase structure and the dense matter equation of state.

Electromagnetic probes, in particular measurements of dilepton emission, is a promising tool for studying the properties of dense and hot matter~\cite{Shuryak:1978ij,McLerran:1984ay,Bratkovskaya:1996qe,Rapp:1999ej}.
They are of particular elegance compared to hadronic observables, as the hot hadronic medium is virtually transparent for electroweak interactions~\cite{Rapp:2014hha}.
Hence, dileptons reflect the properties of the medium at the time and place of their emission.
Inclusive dilepton spectra represent cumulative properties of the system at different times in its evolution.
In collisions that create baryon rich matter, one expects great sensitivity of the electromagnetic probes to the hadronic equation of state and a possible phase transition~\cite{Santini:2011zw,Rapp:2014hha,Li:2016uvu}.
At collisions in the vicinity of the phase transition, the strongest effect is expected from the slowed expansion which results in an extended lifetime which consequently leads to a longer emission \cite{McLerran:1984ay,Heinz:1990jw,Rischke:1996em}.
Dynamical simulations have shown this effect qualitatively for dileptons \cite{Li:2016uvu} and recently a full 3+1D ideal fluid dynamic simulation has confirmed this effect in quantitative more realistic simulation which predicted an increased emission of dileptons up to a factor of 2 \cite{Seck:2020qbx}.
On the other hand, it was discussed that the existence of a partonic phase can also lead to an enhancement of real photons at SPS and FAIR energies \cite{Bauchle:2010yq}.

The present work shows how the inclusion of a phase transition in dense QCD matter influences the total measured dilepton yield in the SIS18--SIS100 beam energy range.
To make quantitative and realistic predictions we use a microscopic transport model which naturally includes viscosity and finite size effects.
In the following, it is shown how the well tested UrQMD transport code can be modified to include a density dependent first-order phase transition based on a many-body interaction potential.
Predictions for possible dilepton measurements, sensitive to the phase transition and in particular on the extended lifetime in a realistic scenario, are then presented. 

\section{Dynamical description of heavy ion collisions with a phase transition}

An estimate of the dilepton emission from the rapidly expanding fireball of a relativistic HIC requires a detailed microscopic description of said evolution.
For this we will employ the well known and tested UrQMD hadronic transport model in its most recent version (v3.5)~\cite{Bass:1998ca,Bleicher:1999xi,Bleicher:2022kcu}.
It is based on the explicit propagation of hadrons in phase space, their two-body scatterings, and decays of unstable particles.
The imaginary part of the interactions is included through binary elastic and inelastic collisions (the latter leading mainly to resonance excitations and decays at the energies considered in the work, or color flux-tube formation and their fragmentation at higher energies).
The real part of the interaction potential is implemented via a density dependent single particle potential $U(n_B)$.
Using the single particle potential, the mean field potential energy $V(n_B)$ at density $n_B$ can be calculated to solve the classical equations of motions.
Traditionally, the field energy per baryon is computed using a simple Skyrme potential~\cite{Bass:1998ca, PhysRevC.101.035205}.
However, recently it was shown how a more sophisticated relativistic Chiral Mean Field model (CMF) can also be incorporated~\cite{OmanaKuttan:2022the}.
In the following, the CMF EoS in UrQMD will be used, as well as two modifications of the CMF equation of state (EoS) that produce phase transitions at high baryon densities, with a focus on studying the effect of the phase transitions on the dilepton emission in the SIS100 energy regime in a consistent approach.
For previous dilepton studies within UrQMD we refer the reader to \cite{Vogel:2007yu,Schmidt:2008hm,Bratkovskaya:2013vx}.

\subsection{The Equations of State}
The treatment of the density dependent potential in UrQMD as well as the construction and motivation for the phase transitions used in the following has been described in more detail in \cite{Steinheimer:2022gqb}.
In the following we will simply give a short summary and refer the interested reader to the above paper.
As a baseline for the density dependent EoS
we use the CMF model in its most recent version~\cite{Motornenko:2019arp}.
The CMF model incorporates a realistic description of nuclear matter with a nuclear incompressibility  of $K_0=267$ MeV, chiral symmetry breaking in the hadronic and quark sectors, as well as an effective deconfinement transition.
Recently it was shown how the CMF EoS can be effectively implemented into the QMD equations of motion of the UrQMD model~\cite{OmanaKuttan:2022the}: 
\begin{eqnarray} \label{QMD_eq}
\dot{\textbf{p}}_{i} & = &-\frac{\partial \rm H }{\partial \textbf{r}_{i}} =  -  \frac{ \partial \mathrm{V} }{\partial \textbf{r}_{i}} \\
  & = & - \left(\frac{ \partial V_i }{\partial n_i}\cdot \frac{\partial n_i}{\partial \textbf{r}_{i}} \right)-\left( \sum_{j \ne i} \frac{\partial V_j}{ \partial n_j} \cdot \frac{\partial n_j}{\partial \textbf{r}_{i}}\right)~, \nonumber
\end{eqnarray}
where $\mathrm{H}$ is the total Hamiltonian and ${\mathrm{V}=\sum_j V[n_B(r_j)]}$ is the total potential energy of the system\footnote{Here momentum dependence is neglected.}.
The potential derivative in equation (\ref{QMD_eq}) is calculated by summing over all possible baryon pairs $j \neq i$.
$V_i$ corresponds to the potential energy of a baryon at position $r_i$ and the local interaction density $n_k$ at position $r_k$ is calculated assuming that each particle can be treated as a Gaussian wave packet~\cite{Aichelin:1991xy,Bass:1998ca}.

\begin{figure}[t]
  \includegraphics[width=.49\textwidth]{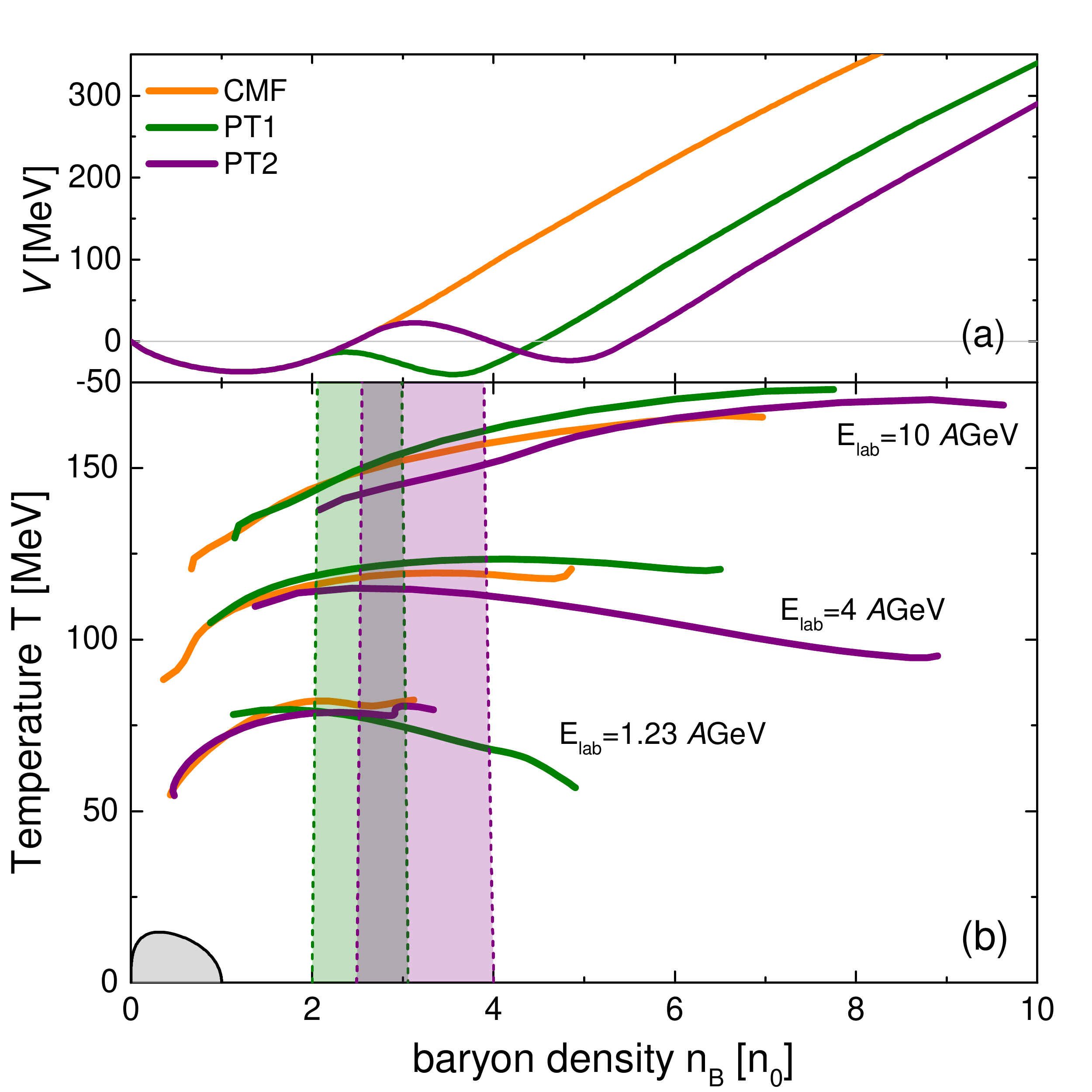}
  \caption{\label{pd} Panel (a): The effective potential of baryons as function of the baryon density $n_B$ for all three equations of state. 
  The standard CMF version monotonously increases at higher densities, while the PT1 and PT2 scenarios incorporate additional phase transitions which are represented by the secondary minima.
  Panel (b): The event averaged trajectories of the central cell are shown as calculated using the coarse graining method for different beam energies. The spinodal regions of PT1 and PT2 in the $n_B$-$T$ phase diagram are indicated by shaded regions.
  }
  \end{figure}

In addition to the derivative of the mean field potential energy per baryon $\partial V(n_B)/\partial n_B$ that is used as input to UrQMD, only the local densities and gradients needed to be calculated to solve the equations of motion for each baryon.
The baseline CMF mean field energy per baryon $V(n_B)$ as a function of the baryon density is shown as an orange line in the upper panel of Fig.~\ref{pd}.
The minimum around saturation density corresponds to the nuclear liquid-gas transition, at higher density the potential monotonously increases due to the 
presence of repulsive interactions and
absence of any other phase transitions.

A simple augmentation is used in order to implement a phase transition in the CMF model \cite{Steinheimer:2022gqb}.
To provide for another meta-stable state in the mean field energy per baryon at large densities (in addition to the bound state from the nuclear liquid-gas transition), the original potential of CMF is cut at density $n_B^{cut}=2.4\,n_0$ for PT1 and $3.0\,n_0$ for PT2.
$V(n_B)$ for $n_B>n_B^{cut}$ is then shifted by $\Delta n_B= 2.0\,n_0$ and $3.0\,n_0$, respectively. 

 \begin{figure*}[t]
  \includegraphics[width=.99\textwidth]{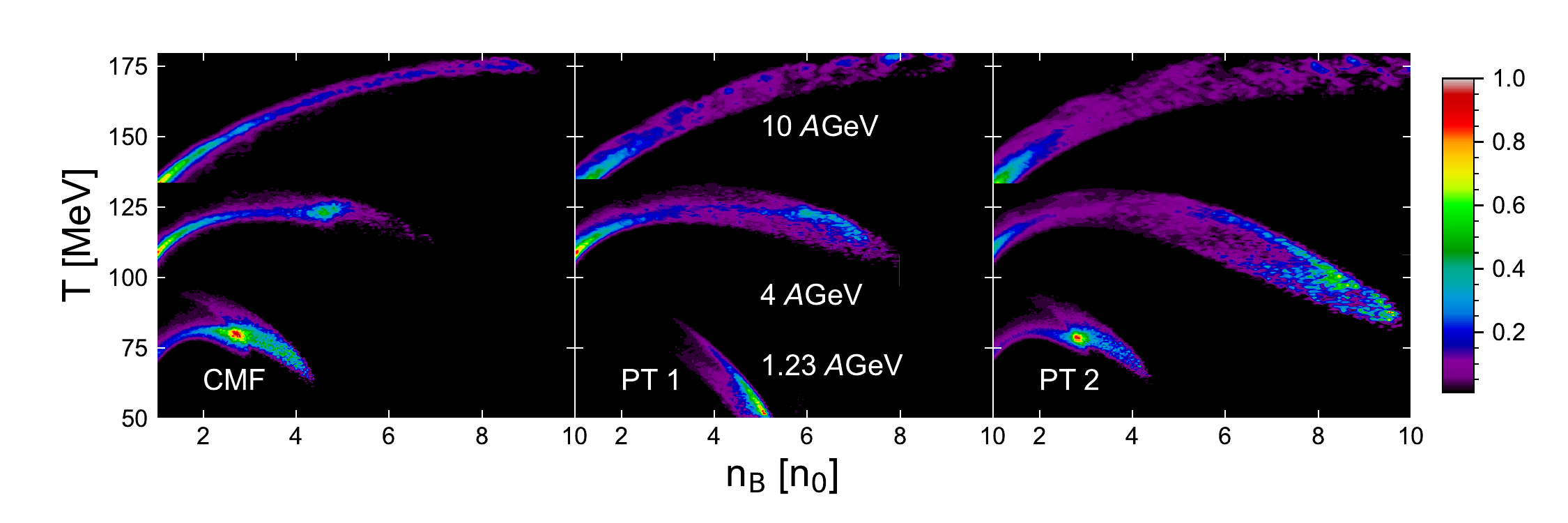}
  \caption{\label{emmision} Time-integrated dilepton emission $1/N_{l\bar l}^{\rm max} d^2N_{l\bar l} / dT\,dn_B $, normalized to its maximum value $N_{l\bar l}^{\rm max}$, for central Au-Au collisions and the three different equations of state and three different beam energies. The results are calculated using the coarse graining approach. The emission is shown as a function of temperature and density, highlighting the regions in the phase diagram which contribute most to the total dilepton excess. Clear differences between the equations of state can be observed, especially if the system expands through or near the expected phase transition.
  }
  \end{figure*} 

The mean field energy between $n_B^{cut}<n_B<n_B^{cut}+\Delta n_B$ is then interpolated by a third order polynomial in order to create a second minimum in energy per particle $V(n_B)$ and ensure that also its derivative is a continuous function.
There are infinite ways to create such a construction, each leading to different properties of the so constructed transition.
Other methods can be employed to construct high density phase transitions, for example using a vector density functional~\cite{Sorensen:2020ygf}.
The present work is interested in studying general signatures of a significant phase transition in the SIS100 energy regime, so we chose to add a significant dip in the potential for densities that are low enough to be reached by heavy-ion collisions at the SIS18/SIS100 accelerators. 
Since this procedure modifies the CMF EoS only at high densities it leaves the low-density description consistent with nuclear matter properties and lattice QCD constraints.

The resulting average field energy per baryon used for the CMF and PT1 as well as  PT2 equations of state are shown as orange, green and purple lines in panel (a) of Fig.~\ref{pd}.
Here, the explicitly visible additional minima of the field energy lead to the presence of two coexisting different-density states.
One should note that this kind of phase transition does not come with a change in the underlying degrees of freedom, since the basic degrees of freedom of the UrQMD transport model are hadrons. Therefore, any effects which would specifically depend on the deconfinement aspects of the transition will not be included in the transport simulation, however, can be included in the coarse graining procedure.
  
To have a better understanding of the expected influence of the modified potential on the phase structure the phase diagrams of the two CMF-PT models, in the thermodynamic limit, are constructed and shown in panel (b) of figure \ref{pd}.
The spinodal lines~\cite{PhysRevC.101.035205}, boundaries of regions where matter becomes mechanically unstable, are defined by the densities at which the isothermal speed of sound becomes imaginary.
The pressure as well as the chemical potential at given ($n_B,T$) can be calculated using the single particle energy:
\begin{eqnarray}
U(n_B)&=&\frac{\partial (n_B  \cdot V(n_B))}{\partial n_B}~,  \\
 \mu(n_B) &=& \mu_{\rm id}(n_B) - U(n_B)~, \\
 P(n_B,T) & = & P_{\rm id}(n_B,T) + \int_0^{n_B} n' \frac{\partial U(n')}{\partial n'} dn'\,.
\end{eqnarray}
where $ P_{\rm id}(n_B,T)$ and $\mu_{\rm id}(n_B)$ are the pressure and chemical potential of an ideal hadron resonance gas.
Having calculated the pressure one can define the spinodal lines which are shown as dashed lines in figure~\ref{pd}.
The shaded areas are the spinodal regions of the corresponding EoS.
Note that the phase transitions implemented in this work are rather strong and since we did not consider any explicit temperature dependence of the potential, there is no critical endpoint present at any finite temperature in both cases. 

Panel (b) of Fig.~\ref{pd} also includes the averaged trajectories of the central cell for central collisions at different beam energies, obtained by the coarse graining method described below.
In all three cases, the nature of the nuclear liquid-gas transition is not changed and is shown as the gray shaded area. 
The trajectories are shown only after the point of highest compression is reached and the maximal density achieved during the compression of the colliding nuclei is a direct result of the stiffness of the EoS.
At the lowest beam energies the maximally reached compression in the CMF and PT2 cases is similar, whereas in the PT1 case one reaches much larger densities due to the phase separation.
At the highest energy, the effect of the transition for PT1 is already much less significant, since the maximum compression is beyond the spinodal region.
Therefore, we would expect to see effects of the phase transition for PT1 already at the lowest beam energy, while effects for PT2 should show up for $E_{\mathrm{lab}}=2$ $A$GeV.
In addition the softening will also lead to a prolonged lifetime of the system and thus larger dilepton emission as has been shown recently in a fluid dynamic simulation \cite{Seck:2020qbx}.
In the following, it is studied whether such a signal survives the bulk evolution in a fully microscopic non-equilibrium model.

\subsection{Coarse Graining}
To calculate the dilepton emission during the collision, a coarse graining approach will be employed.
This approach was used in \cite{Bravina:1998pi,Bravina:1999dh, Huovinen:2002im,Endres:2013daa,Endres:2014zua,Endres:2015fna,Endres:2015egk,Galatyuk:2015pkq,Endres:2016tkg,Reichert:2020yhx,Reichert:2020oes} to extract the properties of the medium from microscopic transport simulations.
For this procedure, 50000 events of the most central ($b<2$~fm) Au-Au collisions are simulated for several beam energies, using either of the three equations of state introduced above.
To calculate the medium properties, an ensemble average of the local energy- and baryon-densities is calculated on a space-time grid with $\Delta x,\Delta y,\Delta z, \Delta ct=0.5$~fm.
The grid constitutes $(100)^4$ cells spanning $-25<x$ [fm] $<25$ in space and $0<t$ [fm/$c$] $<50$ in time.
This allows one to incorporate most (if not all) of the system created in Au-Au collisions at energies  $E_{\rm lab}=1-10$ $A$GeV. 

The medium properties, i.e. the baryon density and temperature, can be extracted from the CMF EoS only in the local rest frame of each gridpoint.
Therefore, the local flow velocity $u=\gamma(1,\vec{v})$ has to be calculated using the ideal fluid equilibrium stress-energy tensor:
\eq{
T^{\mu\nu}=(\varepsilon+P)u^{\mu}u^{\nu}-Pg^{\mu\nu},
}
where $\varepsilon, P$ are the energy density and pressure in the co-moving frame.
Using the energy-, momentum-, and baryon-densities extracted from the coarse grained UrQMD simulations and the pressure from the equation of state, the local rest frame properties $n_B$ and $T$ can then be found, using a fixed-point iterative method used also in ideal fluid dynamical simulations~\cite{Rischke:1995ir}.\footnote{Note, that in this method the extracted temperature depends on the equation of state used. This can lead to different results of the underlying degrees of freedom are not the same as in the microscopic simulation.}

Note that we employ the coarse graining method for the early stage of the collision when the system is far away from thermal equilibrium.
Though the relaxation time is predicted to be smaller than the interpenetration time~\cite{Seck:2020qbx}, the emission of dileptons in the very early stages of the collisions is often not considered because of this uncertainty.
In addition to extracting the local density and temperature, the coarse graining procedure allows to determine the local densities of $\pi,\,\rho-$mesons and $\Delta-$baryons which can be used to check whether system is in a local chemical equilibrium.

It was found that the pion multiplicity $n_{\pi}$ quickly exceeds the statistical model prediction of ideal pion gas at a cell temperature $n^{\rm id}_{\pi}(T)$ by a factor 
\eq{
\lambda_{\pi}=\frac{n_{\pi}}{n_{\pi}^{\rm id}(T)}~,
}
which can be referred to as pion fugacity.
To take into account the pion number excess the values of $\lambda_{\pi}$ are used in calculations of the dilepton emissivity.

\subsection{Dilepton emissivity}
Once the systems space-time evolution is known from the coarse graining, the emitted dileptons and their spectra as function of the pair invariant mass $M$ and pair momentum $k$ can be inferred from the emission rate:
\begin{eqnarray}
\frac{d^8N}{d^4xd^4k}&=&  - \lambda_{\pi}^{1.3}\frac{\alpha^2}{\pi^3 M^2}f^{BE}(k^0,T)\frac{1}{3}g^{\mu\nu} \nonumber \\
 \quad &\times & \operatorname{Im}\left[\Pi^{\mu\nu}_{EM}(M,k,n_B,T)
 \right]~,  
\end{eqnarray}
where $\alpha$ is the fine structure constant and the power of 1.3 for the pion fugacity comes from the fact that not only pions and the $\rho$ meson, but also direct decays from other resonances which contain one or more pions, will contribute to the dilepton emissivity. 
The in-medium properties enter through the spectral function $\Pi^{\mu\nu}_{EM}(M,k,n_B,T)$, where $n_B,\,T$ are the co-moving frame baryon density and temperature.
In the present work we use the most recent version of the Rapp-Wambach-van-Hees spectral function  as introduced in \cite{Rapp:1999us,vanHees:2007th,Rapp:2013nxa}.
This spectral function describes the EM properties of hot and dense hadronic matter calculated from the in-medium $\rho$-, $\omega$- and $\phi$ meson spectral functions from hadronic many-body theory~\cite{Rapp:1999ej}, supplemented by a 4-pion continuum with chiral mixing at masses above 1~GeV.
The calculated dilepton emission spectra can then be directly compared to the experimentally measured dilepton excess where the initial- (NN Bremsstrahlung, Drell-Yan) and final state decays are already subtracted.
It should also be noted that this in-medium rate becomes more similar to a pure quark rate as the temperature or density is increased, which is a consequence of the quark-hadron duality near the deconfinement transition.
This also means that the spectral function is in principle also valid for densities above the deconfinemnt transition and thus allows us to make realistic predictions on the effect of the phase transition.
Further explicit effects on the spectral function, e.g., from a chiral critical point \cite{Tripolt:2018jre}, are not part of this work.

  \begin{figure}[t]
  \includegraphics[width=.49\textwidth]{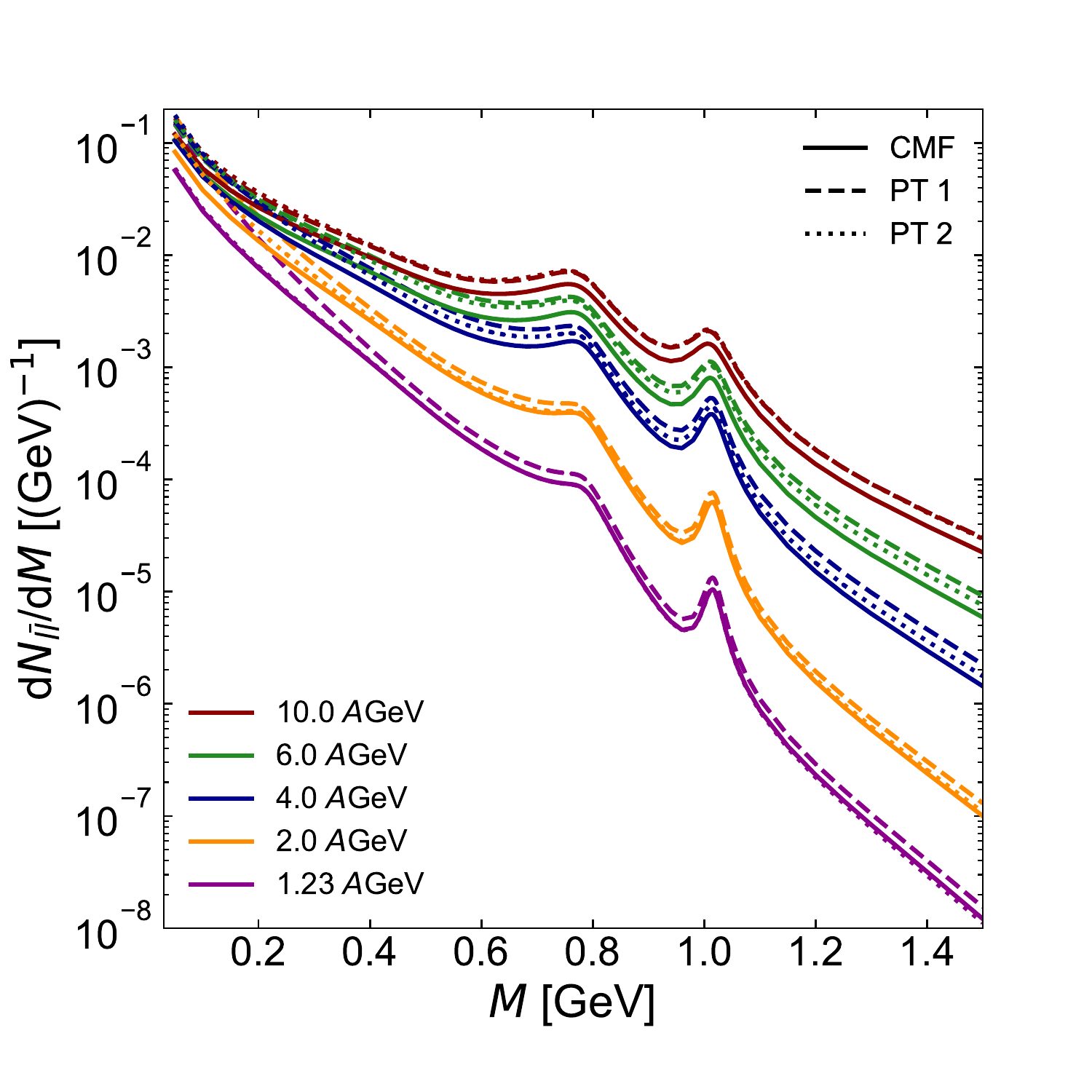}
  \caption{\label{spectra}Invariant mass dilepton spectra in central Au-Au collisions calculated using the CMF and the two CMF-PT equations of state for energies available at SIS18 and at future SIS100 accelerator. Changes in the hadronic in-medium spectral function can be observed from 1.23 to 10 $A$GeV.}
  \end{figure}

\begin{figure}[t]
\includegraphics[width=.49\textwidth]{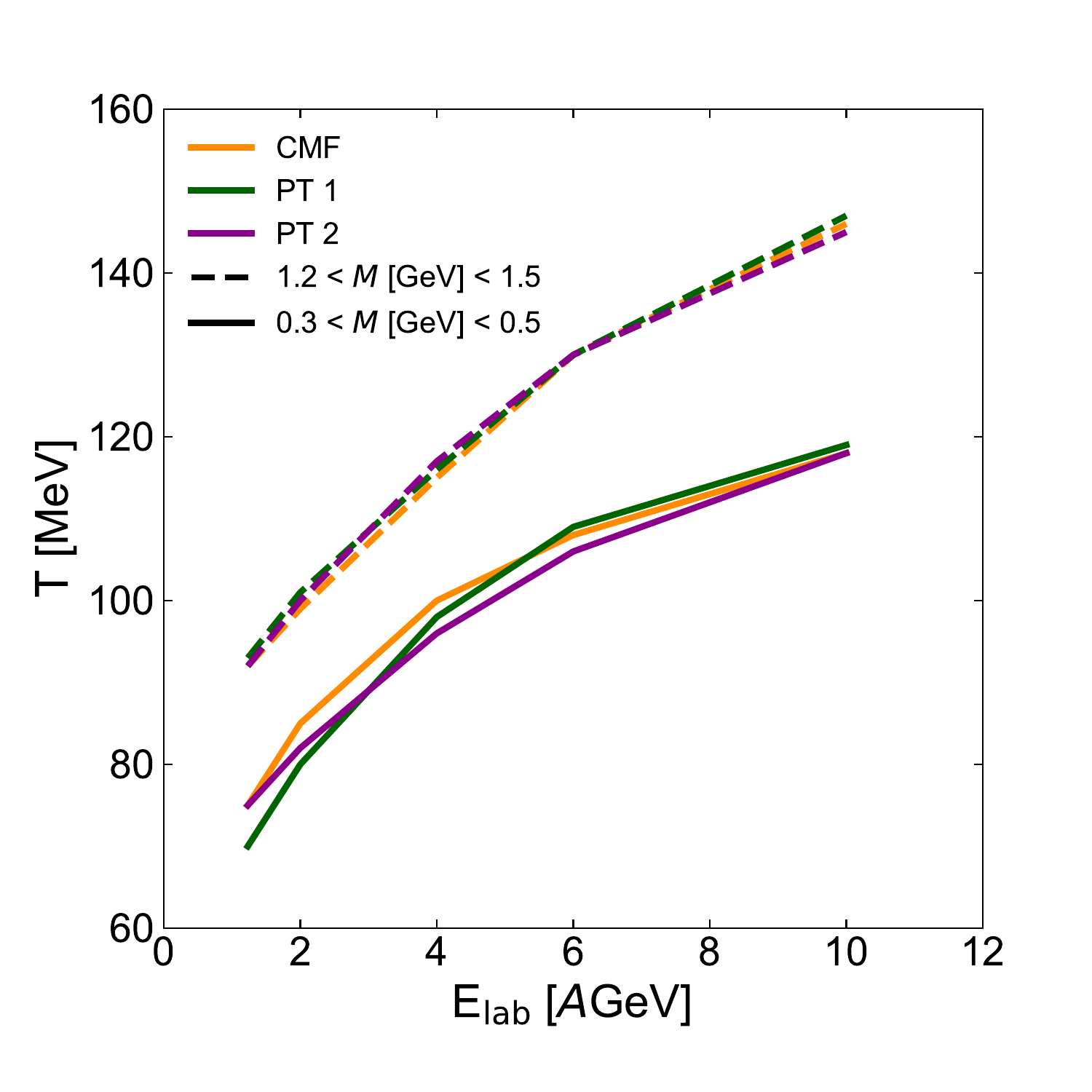}
\caption{\label{exit-func-t} Excitation function of the dilepton temperature, extracted through exponential fits to the mid-rapidity dilepton invariant mass spectrum in two different mass ranges. The low mass range shows a smaller temperature and a stronger sensitivity to the existence of a phase transition.
}
\end{figure}

To understand and interpret the resulting dilepton spectra and yields it is instructive to look first at the regions in the phase diagram where the dileptons are emitted in the three different scenarios. Figure~\ref{emmision} shows the normalized emission rate as function of the baryon density and temperature for the CMF, as well as PT1 and PT2 EoS (from left to right).
For each EoS three separate beam energies are presented, $E_{\mathrm{lab}}= 1.23$, 4 and 10~$A$GeV, from bottom to top.
One can clearly observe that, once the system undergoes the phase transition and/or softening of the EoS, the emission at high baryon densities increases.
This occurs for PT1 already at 1.23~$A$GeV and for PT2 at 4~$A$GeV.
Therefore, for the case of a phase transition, we would expect an increased dilepton emission and possible increase of the extracted dilepton temperature at those beam energies.

\section{Results}
The time and momentum integrated (excess) invariant mass spectra of dileptons for central Au-Au collisions at five different beam energies, calculated with the hadronic in-medium spectral function, is presented in Fig.~\ref{spectra}.
Here, the CMF EoS (solid lines) is compared to the two phase transition scenarios PT1 (dashed lines) and PT2 (dotted lines). 

In addition to the peaks for the $\omega$ and $\phi$ mesons, the spectrum is very flat and no explicit $\rho$ peak can be observed.
A slight broadening of the two vector mesons is observed with the increase of beam energy, and thus the increase of the systems density. Importantly, a systematic enhancement of the spectra as a result of the phase transition is observed. 

To get a better quantitative understanding of the effect of the phase transitions, the inverse slope as well as the integrated yields can be useful tools.
Thus, figure \ref{exit-func-t} shows the dilepton temperature, extracted from an exponential fit to the invariant mass distribution in two mass ranges, $0.3<M<0.5$~GeV (solid lines) and $1.2<M<1.5$~GeV (dashed lines) as functions of the beam energy. These mass ranges have been selected to avoid any contamination of the $\omega$ and $\phi$ in the fit.
As expected the low mass range shows a smaller temperature than the intermediate mass range and the extracted temperature is increasing monotonically with the beam energy. 
The effects of the phase transition are essentially not visible in the intermediate mass range. However, the low mass range is more sensitive. A decrease of the temperature is observed for PT1 and PT2, at beam energies when the system passes through the unstable region, which leads the system to emit slightly longer at a lower temperature. The observed effect corresponds to only a few MeV in temperature and thus would need a very precise experimental determination of the temperature to be used as a signal.

\begin{figure}[t]
\includegraphics[width=.5\textwidth]{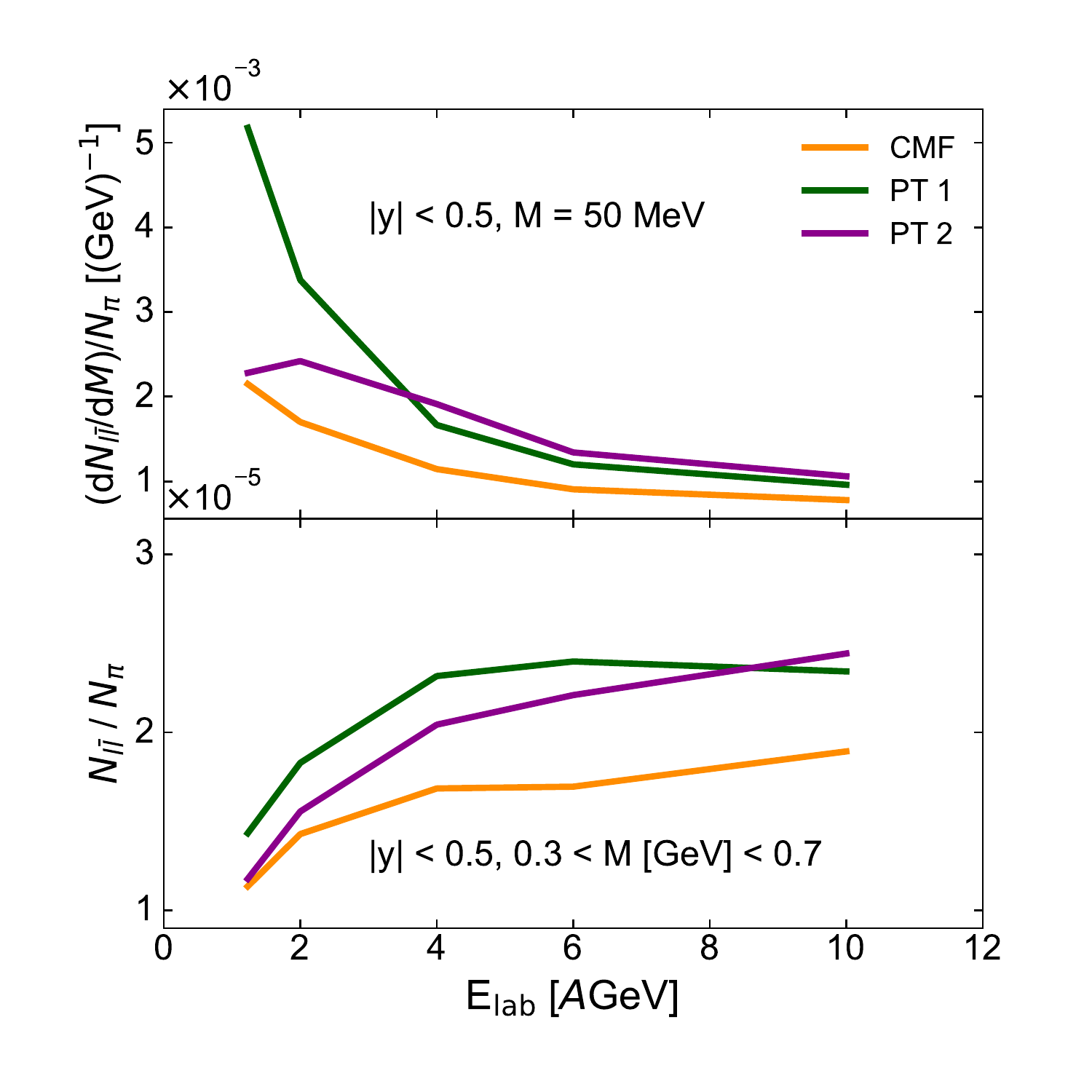}
\caption{\label{exit-func-yield} Upper panel: Dilepton emission rate $dN_{l\overline{l}}/dM$ at $M=50$ MeV, integrated over all momenta at mid-rapidity and normalized by the charged pion yield, as function of the projectile kinetic beam energy. At this low mass the enhancement of the emission rate is very strong if a phase transition is reached. The maximum enhancement occurs at the beam energy when the system enters the phase transition.
Lower panel: Integrated dilepton yield over charged pion number in one unit of rapidity, at mid-rapidity, as functions of the projectile kinetic beam energy. The dileptons are integrated in the low-mass range of the invariant mass distribution. One can observe a clear and significant increase of the dilepton yield once the phase transition (or significant softening) is reached. The maximum in the enhancement is reached only once the initial compression has clearly surpassed the phase transition coexistence densities.
}
\end{figure}

Finally, Fig.~\ref{exit-func-yield} shows the excitation function of the dilepton emission rate (upper panel) as well as the mid-rapidity dilepton yield integrated over the mass range $0.3<M<0.7$~GeV (lower panel), both normalized by the $dN/dy|_{y=0}$ of charged pions, for the three EoS.
The strongest enhancement is observed for the dilepton emission rate $dN_{l\overline{l}}/dM$ at low masses ($M=$~50~MeV).
Here, a factor of 2-3 enhancement can be observed as soon as the created system enters the unstable region of the phase transition. This corresponds also to the beam energy where the effect is maximal, i.e., one can expect a maximum in the excitation function of the rate at that low mass as function of beam energy. 

A clear enhancement of the normalized yield is also visible for both EoS with a phase transition. \footnote{Note, that we have checked that this increase is genuinely due to an increase of the dilepton yield and that the final pion number is almost unaffected by the EoS.}  
While for PT1 the enhancement is already significant at the lowest beam energy, as here already the coexistence phase is reached, for PT2 the enhancement occurs at higher beam energies.
PT1 even shows a maximum enhancement, compared to the CMF EoS, at approximately $E_{\mathrm{lab}}=$~6~$A$GeV, when the systems maximal compression is significantly above the spinodal region and the phase transition effect becomes weaker.
For PT2 the dilepton excess increases up to the highest beam energy under investigation.
However, we also expect to see a maximum at even higher beam energy, especially of a phase transition with a critical endpoint.
Unfortunately, simulating the higher beam energies within the present approach requires a careful treatment of relativistic effects \cite{Ko:1987gp,Ko:1988zz,Blattel:1988zz,Sorge:1989dy,Blaettel:1993uz,Fuchs:1996uv,Buss:2011mx,Nara:2019qfd} which is out of the scope of the current paper.

\section{Conclusions}
It was shown that a first order phase transition in a dynamical non-equilibrium transport description of heavy-ion collisions will lead to a significant enhancement in the measured dilepton yield per charged pion.
Unlike previous studies, in this setup, the whole evolution is treated within a consistent dynamical framework that allows us to describe particle production as well as the dynamical evolution of the system without the introduction of additional parameters.
In addition, since we use a microscopic transport model, effects of finite viscosity and a finite system size are naturally included.
The different magnitude of the signal, as compared to \cite{Seck:2020qbx}, can be understood as a result of the finite viscosity as well as a different treatment of the effective degrees of freedom in the coarse graining procedure as compared to the fully consistent hydro approach.

The predicted enhancement factor of 2-3 for the emission rate at masses of $M\approx 50$~MeV and 1.5 from the integrated yield in the whole low-mass region is still sizeable enough to be measured by the upcoming experiments. 

In addition, a dilepton temperature reduction of about 5~MeV is observed in the low mass range when the system created in the nuclear collision reaches densities in the coexistence region of the transition. 

\begin{acknowledgments}		
The authors thank  L.~Satarov and H.~St\"ocker, R.~Rapp and J.~Stroth for fruitful comments and discussions.
OS acknowledges the scholarship grant from the GET$\_$INvolved Programme of FAIR/GSI.
AM acknowledges the Stern-Gerlach Postdoctoral fellowship of the Stiftung Polytechnische Gesellschaft.
JS thanks the Samson AG and the BMBF through ErUM-Data for funding.
This article is part of a project that has received funding from the European Union’s Horizon 2020 research and innovation programme under grant agreement STRONG – 2020 - No 824093.
This work is supported by the National Academy of Sciences of Ukraine, Grant No. 0122U200259.
VV was supported through the U.S. Department of Energy, Office of Science, Office of Nuclear Physics, under contract number DE-FG02-00ER41132.
MIG acknowledges the support from the Alexander von Humboldt Foundation.
TG acknowledges the support by the State of Hesse within the Research Cluster ELEMENTS (Project ID 500/10.006).
\end{acknowledgments}

\bibliography{refs}

\end{document}